\documentclass{iopart}

\usepackage{graphicx}
\usepackage{dcolumn}
\usepackage{braket}
\usepackage{bm,color}

\begin{document}


\title[Neutrinoless $\beta\beta$ decay
mediated by light and heavy neutrinos: Role of correlations]
{Neutrinoless $\beta\beta$ decay
mediated by the exchange of light and heavy neutrinos:
The role of nuclear structure correlations}

\author{J. Men{\'e}ndez}

\address{Center for Nuclear Study, The University of Tokyo,
113-0033 Tokyo, Japan}
\ead{menendez@cns.s.u-tokyo.ac.jp}

\date{\today}

\begin{abstract}
Neutrinoless $\beta\beta$ decay nuclear matrix elements
calculated with the shell model and energy-density functional theory
typically disagree by more than a factor of two
in the standard scenario of light-neutrino exchange.
In contrast, for a decay mediated by sterile heavy neutrinos
the deviations are reduced to about 50\%,
an uncertainty similar to the one due to short-range effects.
We compare matrix elements
in the light- and heavy-neutrino-exchange channels,
exploring the radial, momentum transfer and angular momentum-parity
matrix element distributions,
and considering transitions that involve
correlated and uncorrelated nuclear states.
We argue that the shorter-range heavy-neutrino exchange
is less sensitive to collective nuclear correlations,
and that discrepancies in matrix elements are mostly due
to the treatment of long-range correlations in many-body calculations.
Our analysis supports previous studies
suggesting that isoscalar pairing correlations, which affect mostly
the longer-range part of the neutrinoless $\beta\beta$ decay operator,
are partially responsible for the differences between nuclear matrix elements
in the standard light-neutrino-exchange mechanism.
\end{abstract}



\noindent{\it Keywords\/}
{~Double-beta decay, nuclear matrix elements, shell model}

\vspace{1.cm}

\section{Introduction}
\label{sec:intro}

Searches for nuclear neutrinoless $\beta\beta$ ($0\nu\beta\beta$) decay
---the process where a nucleus decays into its isobar
with two more protons and two fewer neutrons emitting only two electrons---
aim to establish that neutrinos are its own antiparticles,
a unique property among elementary particles
first proposed by Ettore Majorana eight decades ago~\cite{maj37}.
These experiments are very challenging because
the decay they are looking after is indeed very rare, for at least two reasons.
First, $\beta\beta$ decay implies a second-order process
in the weak interaction, which must be very slow.
Second, the neutrinoless mode violates lepton number
---two leptons are created--- so that
there is a further suppression due to the tiny value of the neutrino masses,
or in general by any small parameter distinctive of the new physics
beyond the Standard Model responsible for the lepton-number-violating decay.
While the lepton-number-conserving $\beta\beta$ decay
with the additional emission of two antineutrinos
has been observed in a dozen nuclei, in some cases
with half-lives as short as $10^{19}$~years~\cite{Barabash15},
present lower limits for the neutrinoless mode
already indicate half-lives longer than about 
$10^{26}$~years for $^{136}$Xe~\cite{KamLAND-Zen16},
and $10^{25}$~years for $^{76}$Ge~\cite{GERDA17} and $^{130}$Te~\cite{CUORE15}.
Ongoing searches and future proposals aim to increase the sensitivity
of $0\nu\beta\beta$ decay experiments by several orders of magnitude
in the next decade~\cite{GomezCadenas11,Cremonesi14,DellOro16}.

The $0\nu\beta\beta$ decay half-life also depends on the
associated nuclear matrix element,
reflecting that the process occurs within atomic nuclei~\cite{Engel17}.
Consequently, matrix elements are crucial to anticipate
the experimental sensitivity needed to expect a detection signal
---within a given model of lepton-number violation---
and likewise, to interpret such a decay signal
by identifying the underlying mechanism responsible for it.
Once the decay channel is known,
information can be gained on the corresponding new-physics model.
For instance, in the standard scenario that
$0\nu\beta\beta$ decay is mediated by
the virtual exchange of the observed light neutrinos,
the nuclear matrix elements would allow
to identify the arrangement of the neutrino masses
---the neutrino hierarchy---
and to extract information on the absolute neutrino mass.

Besides the exchange of the light neutrinos we know to exist,
other mechanisms involving new physics beyond the Standard Model
could trigger the $0\nu\beta\beta$ decay.
These include the existence of lepton and hadron right-handed currents
predicted in left-right symmetric models,
the exchange of supersymmetric particles,
or the emission of Majorons associated with the spontaneous breaking
of the baryon-lepton number symmetry~\cite{ver12}.
One representative example 
recently explored in detail in the literature
is the exchange of sterile heavy neutrinos via left-handed currents.
Sterile neutrinos are predicted in several new-physics models
that incorporate the seesaw mechanism
to explain the tiny masses of the observed light neutrinos~\cite{Blennow10}.
In addition, sterile neutrinos with masses in the eV range
have also been motivated as an explanation to the
anomalies seen in a number of neutrino experiments~\cite{Gariazzo15},
and keV sterile neutrinos have been proposed
as a source of warm dark matter~\cite{keVneutrino17}.
For $0\nu\beta\beta$ decay mediated by sterile-neutrino exchange,
the corresponding nuclear matrix elements can provide access to
the sterile-neutrino mass and to the mixing of electron and sterile neutrinos,
once the $0\nu\beta\beta$ decay half-life has been measured.

It is well known that, at present,
uncertainties in nuclear matrix element calculations
for the light-neutrino-exchange mechanism
\cite{men08,vaq13,sim13,Mustonen13,bar15,yao15,Hyvarinen15,hor16,Iwata16,yao16}
are so large that its capability to give useful guidance
to experiments is in practice limited~\cite{Engel17}.
This shortcoming would similarly prevent from
obtaining reliable information on the neutrino mass and hierarchy
in the event of a $0\nu\beta\beta$ decay signal,
assuming the standard scenario for lepton-number violation.
There are two main sources of uncertainty in matrix element calculations.
First, most of the quoted numbers in the literature ignore a possible
"quenching", or renormalization,
that would reduce the value of the matrix elements up to a factor two or more.
Calculations need such a renormalization factor to correctly reproduce
the half-lives of single-$\beta$ decays
and $\beta\beta$ decays with the emission of antineutrinos,
but its origin is not well understood. Among the most popular explanations are
deficiencies in the many-body approaches, and
missing two-nucleon (or meson-exchange) terms in the transition operator~\cite{Engel17,Pastore17a}.
The consequences for $0\nu\beta\beta$ decay matrix elements
depend on the origin of the "quenching"~\cite{Engel17}.
The additional contribution of the isotensor axial polarizability
to two-nucleon decays was proposed recently~\cite{Shanahan17,Tiburzi17}.
Second, calculations using varied many-body methods severely disagree
in their predictions ---up to a factor two or three depending on the isotope---
illustrating the subtle sensitivity of the $0\nu\beta\beta$ decay operator.
In contrast, the same many-body methods
usually predict rather consistent nuclear structure observables
---such as excitation energy spectra or electromagnetic transitions---
across the nuclear chart.

Much less attention has been devoted to the nuclear matrix element
uncertainties associated with decay mechanisms involving new physics
beyond the Standard Model.
Recently, several calculations have become available
for the case of $0\nu\beta\beta$ decay
mediated by the exchange of sterile heavy neutrinos~
\cite{Blennow10,sim13,Hyvarinen15,hor16,Benes05,Rath13,fae14,Neacsu15,bar15a,Song17,Pastore17b}.
Matrix elements in this channel would be in principle also affected
by the uncertainty associated with "quenching".
The renormalization specific to this mechanism could be unique
because the exchange of virtual heavy neutrinos
is characterized by a shorter-range character and larger momentum transfers,
on which the "quenching" can depend.
This is especially the case if the renormalization
is due to meson-exchange currents~\cite{Menendez11,men16a}.
In addition, heavy-neutrino-exchange matrix elements obtained
with various many-body approaches
can be confronted in the same fashion to the routine comparisons performed
in the light-neutrino-exchange channel.
A combined study of matrix elements in several channels
can bring new insights to understand the inconsistencies
among present calculations.

In this work we compare shell-model nuclear matrix elements
to recently published results
obtained with covariant energy-density functional (EDF) theory~\cite{Song17},
with focus on both light- and heavy-neutrino-exchange mechanisms.
For the heavy-neutrino-exchange channel, the matrix elements
depend much less on the many-body method used
than in the case of exchange of virtual light neutrinos.
In fact, the discrepancies for the heavy-neutrino exchange
are comparable to the uncertainty associated with short-range effects.
The latter uncertainty is much larger for heavy neutrinos
than in the light-neutrino-exchange channel
because in the former case larger momentum transfers become relevant,
and the short-range part of the operator ---at present mostly unexplored---
turns out more important~\cite{Hyvarinen15,fae14,Song17,Rath12}.

Previous studies have pointed out the relevance of nuclear structure
correlations in understanding the deviations between matrix elements
obtained with alternative many-body approaches~\cite{Engel17}.
The objective of this work is to exploit
the distinct nature of the shorter-range heavy-neutrino-exchange
and the longer-range light-neutrino-exchange mechanisms
to investigate the influence
of the leading nuclear structure correlations in each of the two channels.
The study of transitions involving simplified uncorrelated states
can provide an overall picture
of the role played by nuclear correlations~\cite{men14}.
In addition, a separable collective Hamiltonian~\cite{duf96}
can be used to single out particular correlations~\cite{men16}.
Our analysis aims to identify correlations
relevant for the light-neutrino exchange
but not so much for the heavy-neutrino case.
The contrasting comparison of matrix elements in the two decay channels
suggests that the diverse many-body treatment of these correlations
can explain part of the disagreement between calculated nuclear matrix elements.

This paragraph completes the introductory Sec.~\ref{sec:intro}.
The remaining of the article is structured as follows.
In Sec.~\ref{sec:light_heavy}
we introduce the $0\nu\beta\beta$ decay nuclear matrix elements
corresponding to the light- and heavy-neutrino-exchange mechanisms.
We compare the results obtained with the shell model
and covariant EDF theory for each decay channel in Sec.~\ref{sec:nme_compare}.
In addition we investigate the matrix element contributions
in terms of the internucleon distance, the momentum transfer,
and the quantum numbers of the pair of decaying nucleons.
Section~\ref{sec:correlations}
discusses the role of nuclear structure correlations.
We first focus on matrix elements for decays involving
uncorrelated initial and final nuclear states.
Subsequently, we use a a separable collective Hamiltonian to
explore individual nuclear correlations
in terms of their impact on each $0\nu\beta\beta$ decay mechanism,
placing special emphasis on isoscalar pairing correlations.
Section~\ref{sec:summary} summarizes the main findings of the article.

\section{Light- and heavy-neutrino-exchange mechanisms}
\label{sec:light_heavy}

We focus on the $0\nu\beta\beta$ decay mediated
either by the virtual exchange of Standard Model light neutrinos
or by sterile heavy-neutrino exchange.
Here "heavy" is defined with respect to the typical momentum transfer
of the decay, $|\bm{q}|\sim100$~MeV, so that the heavy-neutrino mass
satisfies $m_h\gg |\bm{q}|$.
References~\cite{Blennow10,Benes05,Rath13,fae14,bar15a}
discuss matrix elements for the general case
of $0\nu\beta\beta$ decay mediated by neutrinos of arbitrary mass.
Under the assumptions above the half-life of the process can be written as
\begin{equation}
\label{eq:half-life}
[T^{0\nu}_{1/2}]^{-1}=G^{0\nu}(Q_{\beta\beta},Z)
\left[\left|M^{0\nu}\right|^2  m_{\beta\beta}^2+
 \left|M^{0N}\right|^2  \eta_{\beta\beta}^2\right],
\end{equation}
where $G_{0\nu}(Q_{\beta\beta},Z)$ is a known phase-space factor
which depends on the initial and final nuclear energies and the electron mass,
$Q_{\beta\beta}=E_i-E_f-2m_e$, and on the atomic number $Z$~\cite{kot12}.
$M^{0\nu}$ and $M^{0N}$ are the nuclear matrix elements of the
light- and heavy-neutrino-exchange channels, respectively.
The parameters $m_{\beta\beta}$ and $\eta_{\beta\beta}$
characterise the lepton-number violation, and are given by
\begin{eqnarray}
\label{eq:violationparam}
m_{\beta\beta} &= 1/m_e \sum_{l=\rm{light}} U_{el} m_l, \\
\eta_{\beta\beta} &= m_N \sum_{h=\rm{heavy}} U_{eh} / m_h,
\label{eq:violationparam2}
\end{eqnarray}
with $U_{ej}$ the component of the neutrino mixing matrix
that connects electron flavour
with the light or heavy mass-eigenstate of mass $m_j$.
The electron and nucleon mass $m_N$ are introduced by convention
to make the lepton-number-violation parameters dimensionless.

For both neutrino-exchange channels the matrix elements
can be decomposed according to the three separate
spin structures of the $0\nu\beta\beta$ decay operator:
$S_F=1$, $S_{GT}=\bm{\sigma}_a\cdot\bm{\sigma}_b$
and $S_T=3\bm{\sigma}_j \cdot \hat{\bm{r}}_{ab}
\bm{\sigma}_k \cdot \hat{\bm{r}}_{ab}- \bm{\sigma}_a \cdot \bm{\sigma}_b$,
with $\bm{\sigma}$ the spin operator
and $\hat{\bm{r}}_{ab}$ the unit vector
in the radial direction between the two decaying neutrons.
Therefore the full matrix element consists of three parts:
\begin{equation}
\label{eq:monu}
M = M_{GT}-\frac{g_V^2}{g_A^2}M_{F}+M_T \,,
\end{equation}
with dominant Gamow-Teller ($M_{GT}$), subleading Fermi ($M_F$)
and smaller tensor ($M_T$) components.
It is relevant to note that calculations without isospin conservation
overestimate $M_F$~\cite{men14}.
In this work we take vector and axial couplings
$g_V=1$ and $g_A=1.27$~\cite{pdg14}.
The three components of the matrix element $M_X$
introduced in Eq.~(\ref{eq:monu}) can be defined in a general form
as a function of the mass of the virtual neutrino exchanged in the decay, $m_j$:
\begin{eqnarray}
\label{eq:nme_def}
\lambda_{\beta\beta}\,&M_{X}=\frac{1}{m_e}\sum_{j} U_{ej} m_j \\ \nonumber 
&\times\frac{2R}{\pi} \int_0^\infty \!\!\! \bm{q}^2 \, d|\bm{q}|
\bra{f} \sum_{a,b}\frac{
j_X(|\bm{q}|r_{ab}) h_{X}(|\bm{q}|) S_X}
{\sqrt{\bm{q}^2+m_j^2}\left(\sqrt{\bm{q}^2+m_j^2}+\mu\right)} \tau^+_a \tau^+_b\ket{i}.
\end{eqnarray}
The lepton-number-violating parameter $\lambda_{\beta\beta}$ is either
$m_{\beta\beta}$ or $\eta_{\beta\beta}$,
$j_{GT}=j_{F}=j_0$ and $j_{T}=j_2$ are spherical Bessel functions,
$h_X$ are the neutrino potentials~\cite{Engel17,sim99}
given in~\ref{sec:potentials}, and $R=1.2A^{1/3}$~fm
is introduced by convention to make the matrix elements dimensionless.
A sum is performed over all nucleons in the nucleus,
with the isospin operator $\tau^+_a$
turning nucleon $a$ into a proton if it is originally a neutron.

Equation~(\ref{eq:nme_def}) assumes the closure approximation, so that
only the initial nuclear state $\ket{i}$ with $Z$ protons and $N$ neutrons
and the final nuclear state $\ket{f}$ with $Z+2$ protons and $N-2$ neutrons
are needed to evaluate the nuclear matrix element.
This approximation entails an additional parameter $\mu$,
a representative energy of the states in the intermediate nucleus
with $Z+1$ protons and $N-1$ neutrons that
enter in the second-order perturbation-theory expression of the matrix element
without the closure approximation.
Diverse calculations using
the non-approximated expression~\cite{mut94,sen13,sen14b} 
indicate that with reasonable choices of $\mu\sim10$~MeV
---but with values that slightly depend for each $\beta\beta$ transition---
the closure approximation introduces an error of less than $10\%$
in the light-neutrino-exchange mechanism.

In the limits of light and heavy neutrinos Eq.~(\ref{eq:nme_def}) leads to
\begin{eqnarray}
\label{eq:nme_limit}
\fl
m_{\beta\beta}\,M_{X}^{0\nu}=\left(\frac{1}{m_e} \sum_{j} U_{ej} m_j\right)
\frac{2R}{\pi} \int_0^\infty \!\!\! \bm{q}^2 \, d|\bm{q}|
\bra{f} \sum_{a,b}\frac{
j_X(|\bm{q}|r_{ab}) h_{X}(|\bm{q}|) S_X}
{|\bm{q}|(|\bm{q}|+\mu)} \tau^+_a \tau^+_b\ket{i},
 \\ \fl
\eta_{\beta\beta}\,M_{X}^{0N}=\left(m_N\sum_{j} \frac{U_{ej}}{m_j}\right)
\frac{2R}{\pi} \int_0^\infty \!\!\! \bm{q}^2 \, d|\bm{q}|
\bra{f} \sum_{a,b}\frac{
j_X(|\bm{q}|r_{ab}) h_{X}(|\bm{q}|) S_X}
{m_N m_e} \tau^+_a \tau^+_b\ket{i},
\label{eq:nme_limit2}
\end{eqnarray}
where the parentheses isolate the lepton-number-violation parameters,
see Eqs.~(\ref{eq:violationparam}) and (\ref{eq:violationparam2}).
The rest of the right-hand side
are the expressions for the $0\nu\beta\beta$ decay nuclear matrix elements.

Usually an additional piece is included to account for
short-range correlations missed by the many-body calculation
of the initial and final nuclear states.
Effectively this amounts to replacing the operators defined by
Eqs.~(\ref{eq:nme_limit}) and (\ref{eq:nme_limit2}) by
\begin{equation}
\bra{f}\hat{O}\ket{i} \rightarrow \bra{f}\hat{O}\,g(r_{ab})\ket{i},
\end{equation}
with the function $g(r_{ab})$ parameterized to compensate
for the missing correlations~\cite{sim09}.
For the light-neutrino exchange the impact of the short-range correlations
is almost negligible, but for the heavy-neutrino exchange the
most common parameterizations for $g(r_{ab})$
result in a significant matrix element uncertainty.
This reflects limitations in constraining the short-range physics
in the nuclear many-body calculations.

The heavy-neutrino matrix elements
in Eq.~(\ref{eq:nme_limit2}) involve the four-nucleon diagram,
and in addition one- and two-pion contributions
included in the $h^{AP}$ and $h^{PP}$ terms in~\ref{sec:potentials}.
The range of these pion-pole contributions
in Eq.~(\ref{eq:nme_limit2}) is not very different
to that of the four-nucleon diagram~\cite{Song17}.
A fully consistent approach to heavy-neutrino exchange
is based on chiral effective field theory (EFT)~\cite{pre03},
where one- and two-pion-exchange diagrams are predicted
to contribute at the same chiral order as the four-nucleon diagram,
but are of longer range.
Similar chiral EFT predictions in other decay channels
involving the exchange of a heavy particle are supported by
matrix elements obtained with the
quasiparticle random-phase approximation (QRPA)~\cite{ver12}.
In the chiral EFT framework the hadronic input associated with
the contact and pion-exchange diagrams
needs to be determined with lattice QCD~\cite{Cirigliano17,Nicholson16}.
At the moment, however, the relevant information
is not known for heavy-neutrino exchange.
Nevertheless since the main focus of this work is to compare
matrix elements for long- and short-range contributions,
and to study how nuclear structure correlations
operate for both kinds of terms,
we restrict to heavy-neutrino matrix elements
as defined in Eq.~(\ref{eq:nme_limit2}).
For the same reason, the effects of meson-exchange currents~\cite{Menendez11}
and the isotensor axial polarizability~\cite{Shanahan17,Tiburzi17}
are also neglected.
Still we emphasize that a chiral EFT approach including
these corrections, like indicated in Ref.~\cite{Cirigliano17b},
should be considered in principle
to study heavy-neutrino-exchange nuclear matrix elements.

\section{Nuclear matrix elements
\label{sec:nme_compare}}

\begin{figure}
\includegraphics[width=.76\columnwidth]{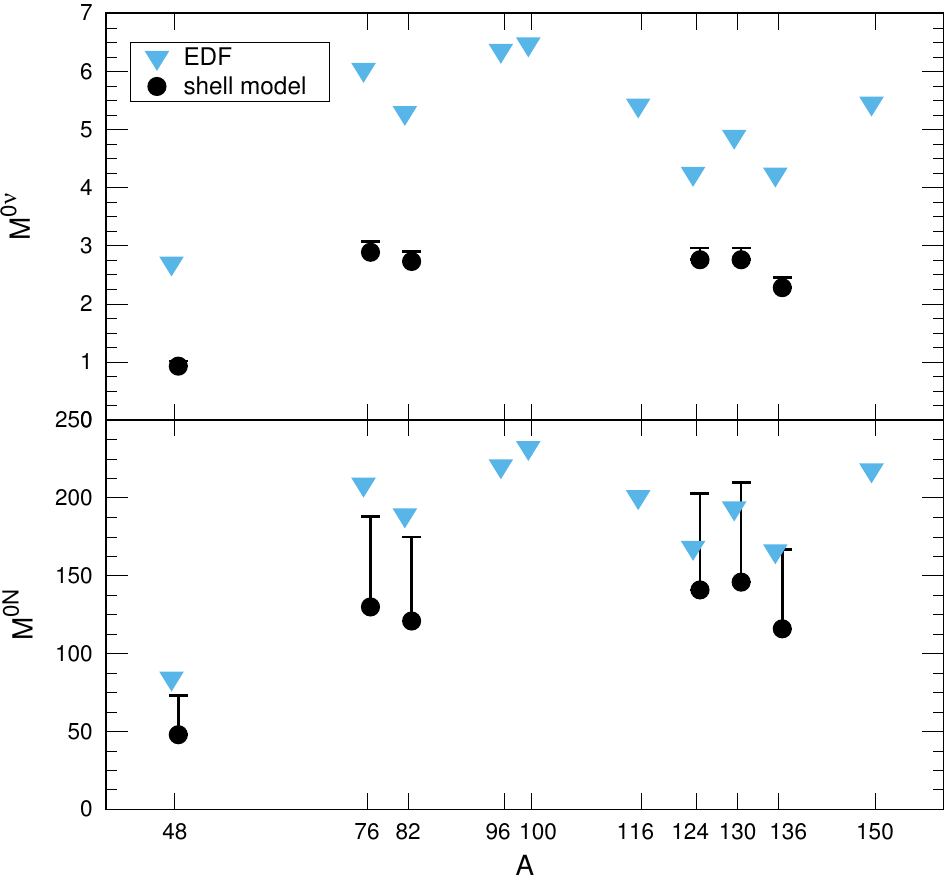}
\caption{Comparison of nuclear matrix elements obtained
with the shell model (black circles) and covariant
energy-density functional (EDF) theory~\cite{Song17} (blue triangles)
for $\beta\beta$ decay emitters with mass number $A$.
Upper panel: Light-neutrino-exchange matrix elements, $M^{0\nu}$.
Lower panel: Heavy-neutrino-exchange matrix elements, $M^{0N}$.
Symbols represent Argonne-type short-range correlations,
and shell-model error bars cover results with CD-Bonn-type correlations as well.
\label{fig:nme_comp}}
\end{figure}

Various many-body methods have been employed to study $0\nu\beta\beta$ decay. 
In the case of light-neutrino exchange it is well known that
different approaches yield matrix elements that vary up to a factor two or more.
The shell model typically gives the smallest matrix elements,
and EDF theory usually the largest ones~\cite{Engel17}.
The upper panel of Fig.~\ref{fig:nme_comp} illustrates this
comparing shell-model matrix elements
with the covariant EDF theory ones of Ref.~\cite{Song17}.
The shell-model values are shown in Table~\ref{tab:nme}.

The shell-model calculations use the same configuration spaces
and interactions as Refs.~\cite{Blennow10,men08,caurier08}: 
For $^{48}$Ca
the neutron and proton single-particle harmonic oscillator orbitals
0$f_{7/2}$, 1$p_{3/2}$, 1$p_{1/2}$ and 0$f_{5/2}$ ---the $pf$ shell---
with the KB3G interaction~\cite{Poves01};
for $^{76}$Ge and $^{82}$Se,
the 1$p_{3/2}$, 1$p_{1/2}$, 0$f_{5/2}$ and 0$g_{9/2}$ orbitals
with the GCN2850 interaction~\cite{men08};
and for $^{124}$Sn, $^{130}$Te and $^{136}$Xe,
the 0$g_{7/2}$, 1$d_{5/2}$,  1$d_{3/2}$, 2$s_{1/2}$ and 0$h_{11/2}$ orbitals
with the GCN5082 interaction~\cite{men08}.
These shell-model interactions are based on
G-matrix interactions~\cite{hjorth-jensen95}
derived from realistic nucleon-nucleon potentials,
with phenomenological modifications, mostly in the monopole part,
to accommodate better agreement with spectroscopic data.
In particular, the GCN2850 and GCN5082 interactions~\cite{Gniady}
describe well spectra
and electromagnetic properties~\cite{sie09,cau10,Klos14,Vietze15}.
All shell-model calculations have been performed
with the $J$-coupled code NATHAN~\cite{cau05},
especially suited for the computation of $0^+$ states.

\begin{table}[b]
\caption{Shell-model nuclear matrix elements for the exchange
of virtual light and heavy neutrinos, $M^{0\nu}$ and $M^{0N}$, respectively.
Left (right) results obtained with
Argonne-type (CD-Bonn-type) short-range correlations.}
\begin{center}
\begin{tabular*}{\columnwidth}{@{\extracolsep{\fill}}ccccccc}
\br
 &
$^{48}$Ca & $^{76}$Ge & $^{82}$Se & $^{124}$Sn & $^{130}$Te & $^{136}$Xe \\
\mr
$M^{0\nu}$ &
$0.93/1.02$ & $2.89/3.07$ & $2.73/2.90$ & $2.76/2.96$ & $2.76/2.96$ & $2.28/2.45$ \\
$M^{0N}$ &
$48/73$ & $130/188$ & $121/175$ & $141/203$ & $146/210$ & $116/167$ \\
\br
\end{tabular*}
\end{center}
\label{tab:nme}
\end{table}

In order to understand the origin of the inconsistency
between matrix element calculations, it is useful to confront the results
for the heavy-neutrino-exchange channel as well,
as in the lower panel of Fig.~\ref{fig:nme_comp}.
A comparison of the upper and lower panels in Fig.~\ref{fig:nme_comp}
highlights that the relative variance between matrix elements 
is much smaller when virtual heavy neutrinos are exchanged
than in the standard scenario of light-neutrino exchange.
Actually we expect the disagreement to be even milder,
because the covariant EDF theory calculations do not preserve
isospin as a good quantum number, and the corresponding matrix elements
---which are larger than the shell model ones---
can be estimated to be about $10\%$ too large
because of this approximation~\cite{men14}. 
Moreover, for heavy neutrinos the discrepancy is comparable
to the uncertainty due to the treatment of short-range correlations
---CD-Bonn- and Argonne-type parameterizations~\cite{sim09} are considered---
represented by the error bars in the shell-model results
in Fig.~\ref{fig:nme_comp}.
What causes the contrasting comparison of matrix elements
in the two channels?

\begin{figure}
\includegraphics[width=\columnwidth]{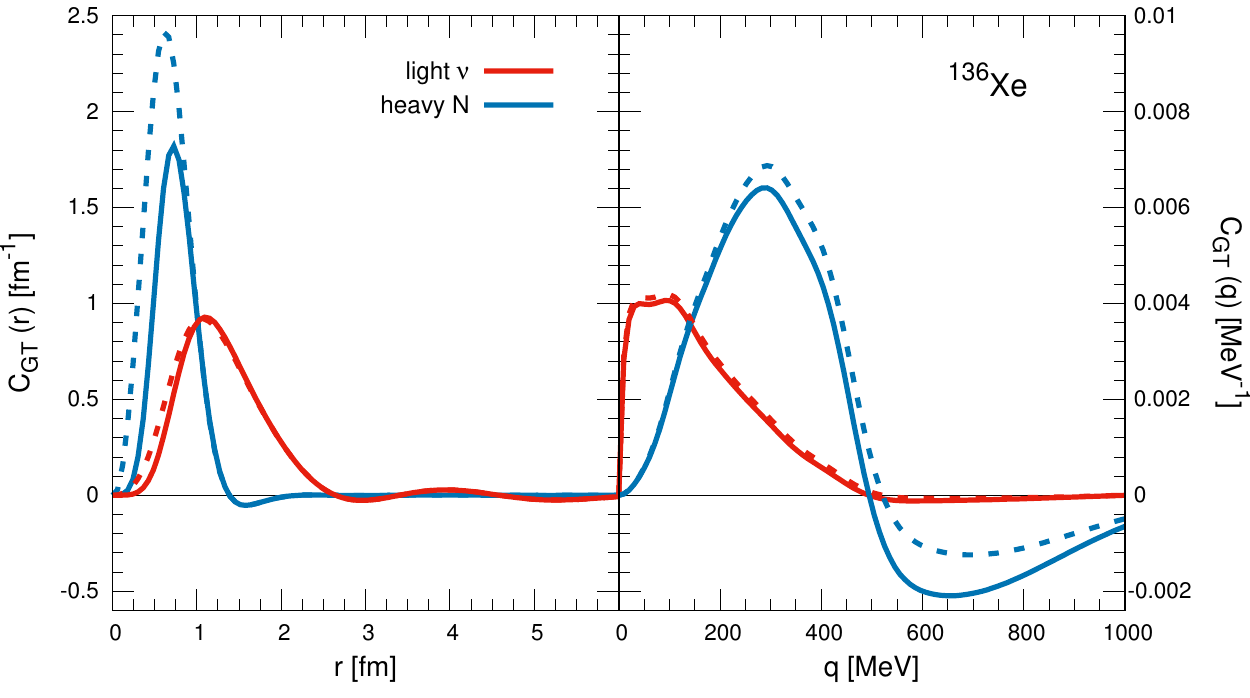}
\caption{Distribution
of the Gamow-Teller matrix element of $^{136}$Xe.
Left panel: Radial distribution.
Right panel: Momentum transfer distribution.
Red (blue) lines show results for the mechanism
that exchanges virtual light (heavy) neutrinos.
Solid (dashed) lines represent Argonne-type (CD-Bonn-type)
short-range correlations. Distributions
normalized with respect to Argonne-type results.
\label{fig:nme_density}}
\end{figure}

To answer this question, it is useful to have in mind
the distinct nature of the two decay mechanisms.
For that purpose, Fig.~\ref{fig:nme_density} compares the corresponding
normalized radial and momentum transfer matrix element distributions,
$C(r_{ab})$ and $C(|\bm{q}|)$, for the representative case of the $^{136}$Xe decay.
The distributions ---without normalization---
satisfy $M=\int C(r_{ab})dr_{ab}$ and $M=\int C(|\bm{q}|)d|\bm{q}|$.
The left panel of Fig.~\ref{fig:nme_density}
shows that even though the light-neutrino exchange is of relatively short-range,
with typical distances between the decaying neutrons
of the order of a couple of fm's, the heavy-neutrino exchange
only allows shorter-range contributions up to $1$~fm.
Alternatively, the right panel of Fig.~\ref{fig:nme_density}
shows that the heavy-neutrino exchange probes momentum transfers
considerably larger than the ones below $|\bm{q}|\approx200$
which are the most relevant for light-neutrino exchange.
Consequently the influence of short-range correlations
in the light-neutrino-exchange channel is minor.
In contrast, for momentum transfers $|\bm{q}|\approx500$~MeV,
only relevant with heavy neutrinos,
short-range correlations play a much more important role.
Here the two prescriptions of short-range correlations
reduce the value of the matrix elements by significantly different amounts,
as shown by the error bars in the lower panel of Fig~\ref{fig:nme_comp}.
The comparison of normalized distributions between the two decay channels
agrees well with the findings of QRPA, EDF theory
and other many-body approaches~\cite{Hyvarinen15,fae14,Song17,Rath12}.

\begin{figure}
\includegraphics[width=\columnwidth]{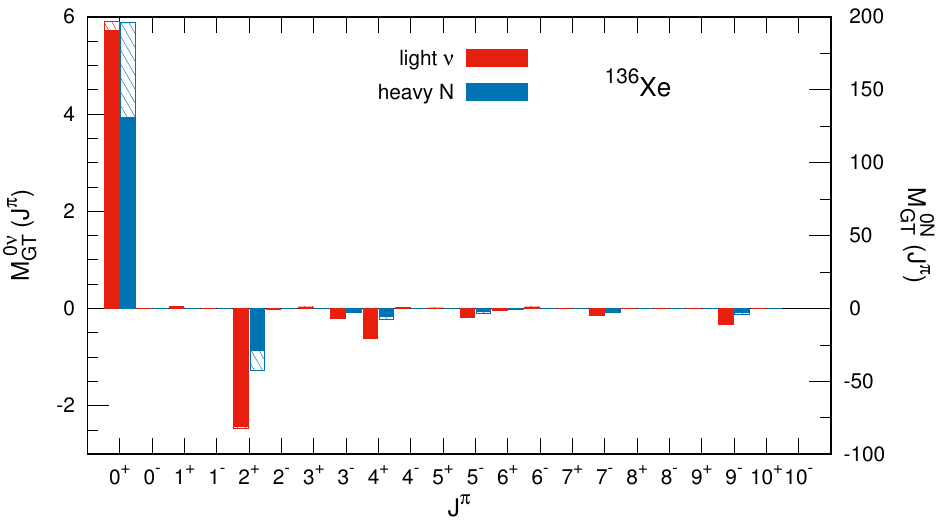}
\caption{Contributions to the Gamow-Teller matrix element of $^{136}$Xe 
from pairs of decaying neutrons with given angular momentum and parity, $J^\pi$.
Red (blue) bars show the results for the exchange of
virtual light (heavy) neutrinos,  $M^{0\nu}_{GT}(J^{\pi})$, left scale
[$M^{0N}_{GT}(J^{\pi})$, right scale].
Solid (dashed) bars correspond to Argonne-type (CD-Bonn-type)
short-range correlations.
\label{fig:nme_jp}}
\end{figure}

We can also analyse the matrix elements in terms of
the contributions from the decaying pair of neutrons
---or the created pair of protons---
coupled to a given angular momentum and parity, $M(J^\pi)$,
with $M=\sum_{J^{\pi}} M(J^{\pi})$.
Figure~\ref{fig:nme_jp}, taking $^{136}$Xe as a representative example,
shows that the decomposition is quite diverse in the two channels.
For the exchange of light neutrinos the total matrix element
results from a significant cancellation of the dominant $0^+$-pair term
by all other contributions, especially those from $2^+$ pairs.
The final result is about $35\%$ of the $0^+$-pair matrix element.
In comparison, for the heavy-neutrino exchange the final result
is about $70\%$ of the $0^+$-pair matrix element.
This term is only modified significantly by $2^+$-pair matrix elements,
as the contributions of pairs coupled
to higher angular momentum are very much suppressed.
These findings are in agreement with Refs.~\cite{hor16,Neacsu15,sen14b}.
The dominance of the $0^+$-pair matrix element
is related to the short-range character of the heavy-neutrino exchange
because for higher angular momenta shorter-range contributions
are much less dominant than for $0^+$ pairs~\cite{sim08}.

The smaller cancellations in the heavy-neutrino exchange matrix elements
suggest that this channel is not as subtle
as the light-neutrino exchange, where the competition
between matrix element contributions needs to be described carefully.
This is consistent with the main finding of Fig.~\ref{fig:nme_comp}:
Matrix elements for the exchange of heavy neutrinos
are similar for the shell model and covariant EDF theory.
Figure~\ref{fig:nme_density} suggests that
most of the matrix element cancellations occur at the level of
longer-range correlations involving nucleons separated by a couple of fm's.

It is useful to explain the matrix element cancellations
invoking an approximate $SU(4)$ symmetry~\cite{men16},
because when the symmetry is exact in the transition operator
and the initial and final nuclear states, the matrix elements vanish.
In this picture, the relatively minor cancellation in heavy-neutrino exchange
shown in Fig.~\ref{fig:nme_jp} occurs
because the very short-range radial part of the $0\nu\beta\beta$ decay operator,
$\sim\delta(r_{ab})$, breaks the $SU(4)$ symmetry
more efficiently than the longer-range $\sim1/r_{ab}$ radial part
corresponding to the light-neutrino exchange.
The $SU(4)$ symmetry can only be fully preserved
for an operator without $r_{ab}$ dependence.

\section{Role of nuclear structure correlations}
\label{sec:correlations}

\subsection{Nuclear matrix elements for uncorrelated nuclear states
\label{sec:correl_range}}

\begin{figure}
\includegraphics[width=.76\columnwidth]{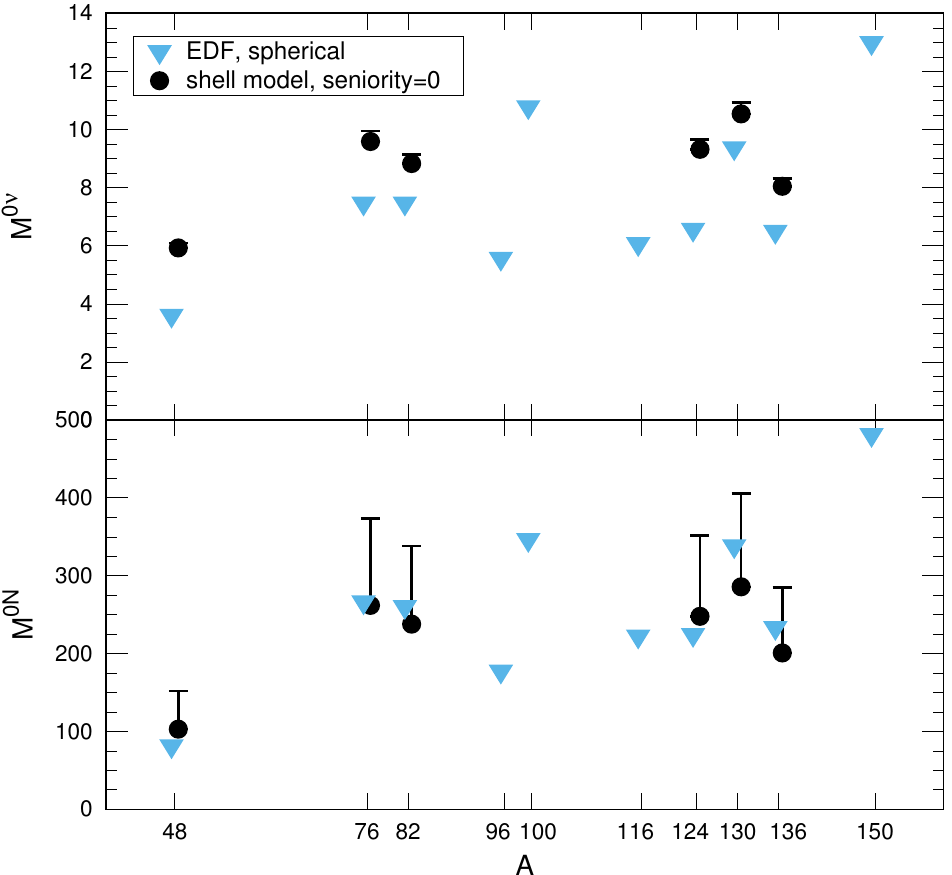}
\caption{Comparison of nuclear matrix elements
obtained with the shell model
restricted to seniority-zero states (black circles),
and covariant energy-density functional (EDF) theory
limited to spherical configurations~\cite{Song17} (blue triangles),
for $\beta\beta$ emitters with mass number $A$.
Upper panel: Light-neutrino-exchange matrix elements, $M^{0\nu}$.
Lower panel: Heavy-neutrino-exchange matrix elements, $M^{0N}$.
Symbols correspond to Argonne-type short-range correlations,
and shell-model error bars cover results with CD-Bonn-type correlations as well.
\label{fig:nme_uncorrel}}
\end{figure}

Given the differences between matrix elements corresponding to
light- and heavy-neutrino exchange, we can expect that
nuclear structure correlations play alternative roles in each channel.
Figure~\ref{fig:nme_uncorrel}
confronts shell-model and covariant EDF theory matrix elements
for calculations restricted to uncorrelated states, for both mechanisms:
Light-neutrino exchange in the upper panel
and heavy-neutrino exchange in the lower panel.
In the shell model the initial and final nuclear states
are limited to configurations formed by neutron-neutron and proton-proton
$0^+$ pairs  ---seniority-zero states---
while for the covariant EDF theory only spherical states are considered.

First, we note that both Figs.~\ref{fig:nme_comp} and \ref{fig:nme_uncorrel}
show a similar matrix element distribution with respect to the mass number
in the upper and lower panels, this is, in the two decay channels.
This suggests a proportionality  between light- and heavy-neutrino-exchange
matrix elements obtained with the same many-body approach,
a dependence first observed in the QRPA~\cite{fae11}.
The non-relativistic EDF theory work Ref.~\cite{Rodriguez13}
found a similar proportionality when comparing matrix elements
for light-neutrino exchange with those corresponding to
the same operator with the radial part replaced by the identity,
a relation studied in more detail in the context of
the correlation between $0\nu\beta\beta$ decay
and double Gamow-Teller transitions~\cite{Shimizu17b}.
These relations suggest that the relative value of the matrix elements
depends mostly on the spin-isospin structure
and correlations of the nuclear states,
and only moderately on the radial part of the operator.
In every channel proposed for $0\nu\beta\beta$ decay
---including those corresponding to
left-right symmetric and supersymmetric models, and Majoron emission---
the dominant spin-isospin structure is a Gamow-Teller-like operator.
Therefore, following Refs.~\cite{fae11,Lisi15} we anticipate that
it will be challenging to identify the underlying decay mechanism
based on a combination of half-life measurements
and matrix element calculations in several isotopes,
as proposed in Refs.~\cite{dep06,geh07}.
Reference~\cite{Cirigliano17b} has very recently obtained a similar conclusion.
In principle small deviations from an exact proportionality would allow one
to distinguish the decay channel~\cite{hor16a},
but a reliable identification demands
more accurate matrix elements than what they are now.

More importantly, it is apparent
in the two panels in Fig.~\ref{fig:nme_uncorrel}
that for uncorrelated states the agreement
between shell-model and covariant EDF theory matrix elements is very good.
Differences are less than $30\%$ in the light-neutrino-exchange channel
---shown in the upper panel---
and only about $10\%$ for the exchange of heavy neutrinos,
as the lower panel shows.
Moreover the uncorrelated shell model matrix elements can be larger
than the EDF theory ones, in contrast to Fig.~\ref{fig:nme_comp}.
This highlights that the nuclear correlations in the the shell model
impact the nuclear matrix elements much more than those included by EDF theory.
The slightly larger shell model values may be related
to a stronger pairing in the shell model interaction
with respect to the pairing strength of the EDF~\cite{men14}.
Note that isospin is not a good quantum number
for the uncorrelated states in any of the two many-body methods,
and this approximation is expected to impact similarly both approaches.

Reference~\cite{men14} found a similar matrix element consistency
for the decay of uncorrelated calcium, titanium and chromium isotopes 
comparing the shell model and non-relativistic EDF theory.
Figure~\ref{fig:nme_uncorrel} expands this agreement
and suggests that it could be extended along the nuclear chart
\footnote{An exception may be heavy systems like $^{150}$Nd
where even calculations with spherical states using
non-relativistic~\cite{rod11} and covariant~\cite{Song17}
EDF theory differ.}.
At this level of uncorrelated states the results only depend on
single-particle degrees of freedom
and like-particle ---isovector--- pairing correlations.
The single-particle structure of non-relativistic EDF theory
in comparison to the shell model has been explored
in Refs.~\cite{Rodriguez16,Rodriguez17}, finding reasonably similar
orbital occupancies obtained with both many-body approaches.
Likewise, isovector pairing is also fairly consistent
in typical shell-model interactions and the Gogny EDF~\cite{cau05}.
Additional like-pairing correlations can be incorporated
by extending the shell-model configuration space~\cite{Iwata16,caurier08a}
or adding isovector pairing fluctuations
to the EDF theory calculation~\cite{vaq13}.
For both methods adding these correlations results in a comparable
enhancement of the matrix element values
of the order of $30\%$ in the $0\nu\beta\beta$ decays studied.
All these similarities suggest that the discrepancy
between shell-model and EDF theory matrix elements
in the standard light-neutrino-exchange scenario is related to
the unequal treatment of additional nuclear structure correlations.

\begin{figure}
\includegraphics[width=\columnwidth]{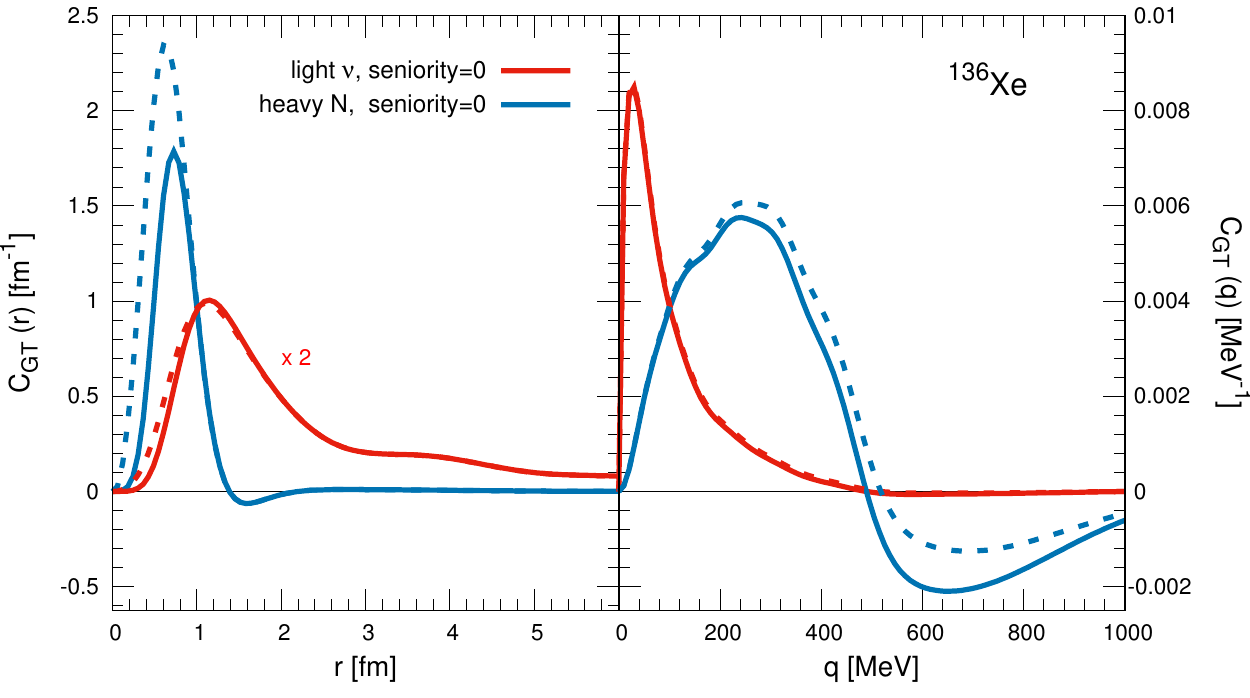}
\caption{Distribution of the 
Gamow-Teller matrix element of $^{136}$Xe
restricted to seniority-zero states.
Left panel: Radial distribution, normalized
to total value 2 for light neutrinos.
Right panel: Momentum transfer distribution.
Red (blue) lines show results for the mechanism
that exchanges light (heavy) neutrinos.
Solid (dashed) lines represent Argonne-type (CD-Bonn-type)
short-range correlations.
The distributions are normalized with respect to Argonne-type results.
\label{fig:nme_density_sen0}}
\end{figure}

The left and right panels of Fig.~\ref{fig:nme_density_sen0}
show the normalized radial and momentum transfer matrix element distributions
$C(r_{ab})$ and $C(|\bm{q}|)$, respectively,
for uncorrelated seniority-zero nuclear states.
The Gamow-Teller $^{136}$Xe matrix element is taken as a representative example.
In the heavy-neutrino-exchange channel the distributions are very similar
to those for fully-correlated states,
see the blue lines in the corresponding panels
in Figs.~\ref{fig:nme_density_sen0} and~\ref{fig:nme_density}.
In contrast, for the exchange of light neutrinos,
a comparison of the red lines in
Figs.~\ref{fig:nme_density_sen0} and~\ref{fig:nme_density}
shows that when seniority-zero states are involved
longer internucleon distances and smaller momentum transfers become relevant.
This suggests that additional nuclear correlations
effectively suppress the longer-range contributions to the matrix element.
Therefore, short-range correlations beyond the seniority-zero level
seem to be relatively under control in what respects $0\nu\beta\beta$ decay
---shell-model and EDF theory matrix elements in the heavy-neutrino channel
agree reasonably well, see the lower panel in Fig.~\ref{fig:nme_comp}.
On the contrary, the discrepancies
in the standard light-neutrino-exchange scenario
---highlighted by the upper panel in Fig.~\ref{fig:nme_comp}---
seem to be caused mainly by nuclear structure correlations
involving nucleons separated by several fm's.
Which are these correlations?

\subsection{Isoscalar pairing correlations}
\label{sec:isoscalar}

Nuclear interactions usually comprise various collective correlations
encoded in the interaction two- and three-nucleon matrix elements.
However, the individual correlations are usually not easy to disentangle.
An exception is the separable collective Hamiltonian $H_\mathrm{coll}$
obtained in Ref.~\cite{duf96}.
This Hamiltonian is valid in the $pf$-shell configuration space
and applicable to nuclei with mass number between $A=40$ and $A\approx60$.
It was constructed to reproduce the most important physics
of the shell-model interaction KB3G.
In fact, for the decay of calcium, titanium and chromium isotopes
$H_\mathrm{coll}$ gives nuclear matrix elements very similar to those of KB3G
in the standard light-neutrino-exchange channel~\cite{men16}.

The separable collective Hamiltonian $H_\mathrm{coll}$
includes the single-particle behaviour
from the monopole part of the KB3G interaction,
and the dominant collective correlations:
\begin{eqnarray}
\label{eq:hsep}
H_\mathrm{coll} = H_M + g^{T=1}
\sum_{n=-1}^1{S}_n^\dag {S}_n
+ g^{T=0} \sum_{m=-1}^1 {P}_m^\dag {P}_m  \nonumber \\
+g_{ph} \sum_{m,n=-1}^1: {\mathcal{F}}^\dag_{mn} {\mathcal{F}}_{mn}: 
+ {\chi} \sum_{\mu=-2}^2 : Q_\mu^\dag Q_\mu :\,.
\end{eqnarray}
Colons indicate normal ordering,
$H_M$ describes the single-particle part and
\begin{eqnarray}
\label{eq:ops-def}
S^\dag_n &= \frac{1}{\sqrt{2}} \sum_{\alpha} \sqrt{2l_\alpha+1} 
\left( a_\alpha^\dag a_\alpha^\dag \right)^{0,0,1}_{0,0,n}\,,  \\
P^\dag_m &= \frac{1}{\sqrt{2}} \sum_{\alpha} \sqrt{2l_\alpha+1} 
\left( a_\alpha^\dag a_\alpha^\dag \right)^{0,1,0}_{0,m,0}\,,  \\
{\cal F}_{mn} &= 2 \sum_{\alpha} \sqrt{2l_\alpha+1} \left( a_\alpha^\dag
\tilde{a}_\alpha \right)^{0,1,1}_{0,m,n}\,, \\
Q_\mu &= \frac{1}{\sqrt{5}} \sum_{\alpha,\beta}\bra{n_\alpha l_\alpha}\!|r^2Y_2/b^2
|\!\ket{n_\beta l_\beta} \left(a_\alpha^\dag \tilde{a}_\beta
\right)^{2,0,0}_{\mu,0,0} \,,
\end{eqnarray} 
where $a^\dag_\alpha$ creates a nucleon in a single-particle orbital with
$n_\alpha$ principal quantum number and $l_\alpha$ orbital angular momentum,
$\tilde{a}_{a}$ destroys a nucleon in the time-reversed orbital,
$Y$ is a spherical harmonic and $b$ the harmonic-oscillator parameter.
Values for the strengths of the isovector pairing, isoscalar pairing,
spin-isospin and quadrupole terms
$g^{T=1}$, $g^{T=0}$, $g_{ph}$ and $\chi$, respectively,
are given in Ref.~\cite{men16}.
In that reference $H_\mathrm{coll}$  was used to isolate the effect
of the individual correlations in the standard $0\nu\beta\beta$ decay channel.
What is the influence of each of these collective correlations
when the transition is mediated by heavy neutrinos?

Isoscalar pairing correlations were shown long time ago 
to be particularly relevant for $\beta\beta$ decay
in the QRPA framework~\cite{vog86,eng88},
an extreme confirmed in improved recent studies~\cite{hin14},
and also with the shell model using $H_{\mathrm{coll}}$~\cite{men16}.
If isoscalar pairing correlations are not taken into account,
the light-neutrino-exchange matrix elements tend to be overestimated.
Here we investigate the impact of isoscalar pairing
in the heavy-neutrino-exchange mechanism.
According to the better agreement between shell model and covariant EDF theory
in this channel, if the discrepancies related to the light-neutrino exchange
are due to isoscalar pairing, the importance of these correlations
would be reduced for the exchange of heavy neutrinos.

\begin{figure}
\includegraphics[width=\columnwidth]{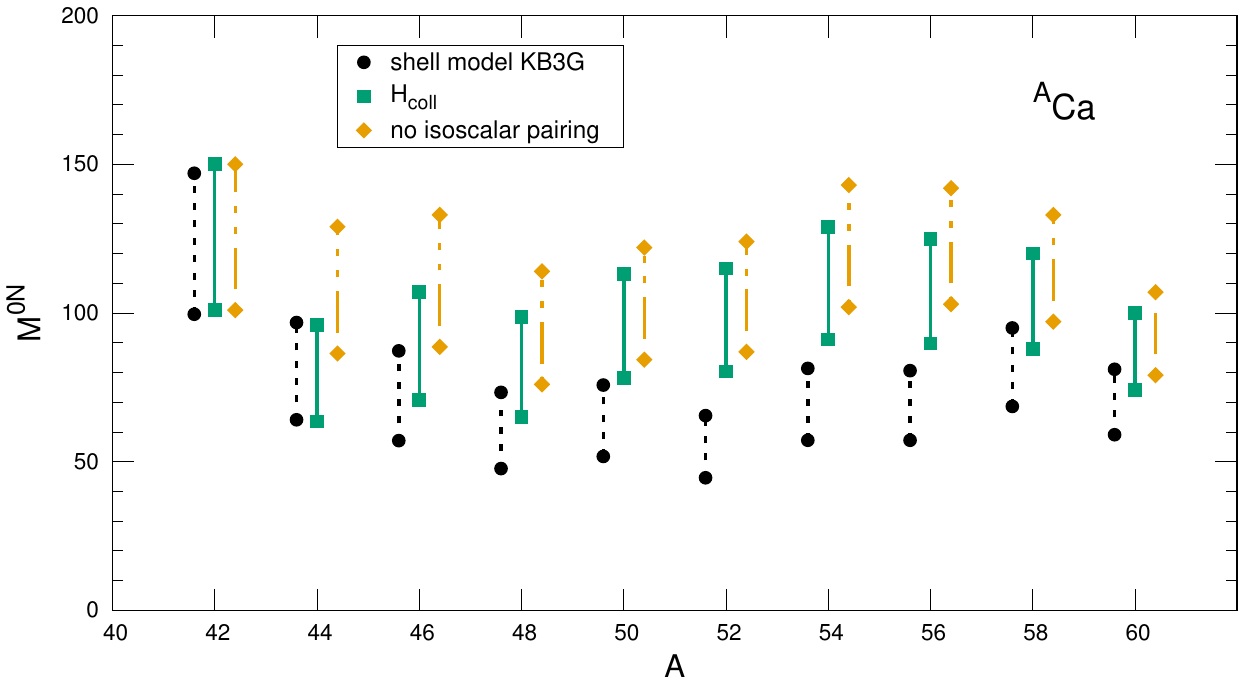}
\caption{Matrix elements for calcium $^A$Ca isotopes
in the heavy-neutrino-exchange channel, $M^{0N}$, 
obtained with the shell-model interaction KB3G (black circles),
the collective Hamiltonian $H_{\mathrm{coll}}$ (green squares),
and $H_{\mathrm{coll}}$ without the isoscalar pairing term (orange diamonds).
Error bars cover results corresponding to Argonne- (lower end)
and CD-Bonn-type (upper end) short-range correlations.}
\label{fig:nme_Hcoll}
\end{figure}

Following a similar strategy to Ref.~\cite{men16} we isolate
the effect of the various terms in $H_{\mathrm{coll}}$
by including them or not in the shell model calculation.
Figure~\ref{fig:nme_Hcoll} shows that isoscalar pairing correlations
indeed play a relatively small role for the heavy-neutrino-exchange mechanism.
Matrix elements for the decay of calcium isotopes
are enhanced by about $10\%-30\%$ when these correlations are omitted,
a correction much less important than
the uncertainty due to short-range correlations
represented by the error bars in Fig.~\ref{fig:nme_Hcoll}.
The enhancement is even smaller than the difference between matrix elements
corresponding to $H_{\mathrm{coll}}$ and to the KB3G interaction.
The impact of isoscalar pairing correlations
in the matrix elements associated with the decay of
titanium and chromium isotopes is very similar.
This minor influence contrasts with the variations observed
in Ref.~\cite{men16} for the $0\nu\beta\beta$ decay of the same nuclei
in the standard light-neutrino-exchange channel.
In that reference, the matrix elements
obtained without isoscalar pairing were found to be a factor of two or three
larger than those corresponding to $H_{\mathrm{coll}}$ or KB3G.

The sensitivity of the heavy-neutrino-exchange matrix elements
to the remaining terms in $H_{\mathrm{coll}}$
---the Landau-Migdal-style spin-isospin and quadrupole correlations---
turns out also very small in all the decays of $pf$-shell nuclei studied.
This is in good agreement with the behaviour
found in Ref.~\cite{men16} in the standard scenario involving light neutrinos.
Furthermore, increasing 
the isovector pairing strength $g^{T=1}$ by $20\%$ 
results in a heavy-neutrino matrix element enhancement of a similar amount, 
which is comparable to the increase found when light neutrinos are exchanged.
This hints that the role of like-particle pairing is similar in both channels.
In conclusion, our analysis based on
the dominant collective nuclear correlations considered in $H_{\mathrm{coll}}$ 
suggests that one of the main differences between
the light- and heavy-neutrino-exchange mechanisms
resides in the importance of isoscalar pairing correlations.

Note that, in addition, quadrupole correlations, which are of long range,
could also cause differences between the two channels.
Quadrupole correlations have been shown to be relevant
in several $0\nu\beta\beta$ decay works, including
shell model and EDF theory studies~\cite{Mustonen13,rod10,men11b,fang11,Song14}.
In general, calculations find reduced matrix elements when
the initial and final nuclear states exhibit different deformations,
a prediction which seems to be in conflict with
measurements of $\beta\beta$ decay with antineutrino emission~\cite{Kidd14}.
Unfortunately, quadrupole correlations play a small role
in the moderately-deformed $pf$-shell nuclei,
and they are relatively unimportant for the decays
shown in Fig.~\ref{fig:nme_Hcoll}~\cite{men16}.
Considering decays involving more deformed systems is necessary
to understand better the influence of quadrupole correlations.

\begin{figure}
\includegraphics[width=\columnwidth]{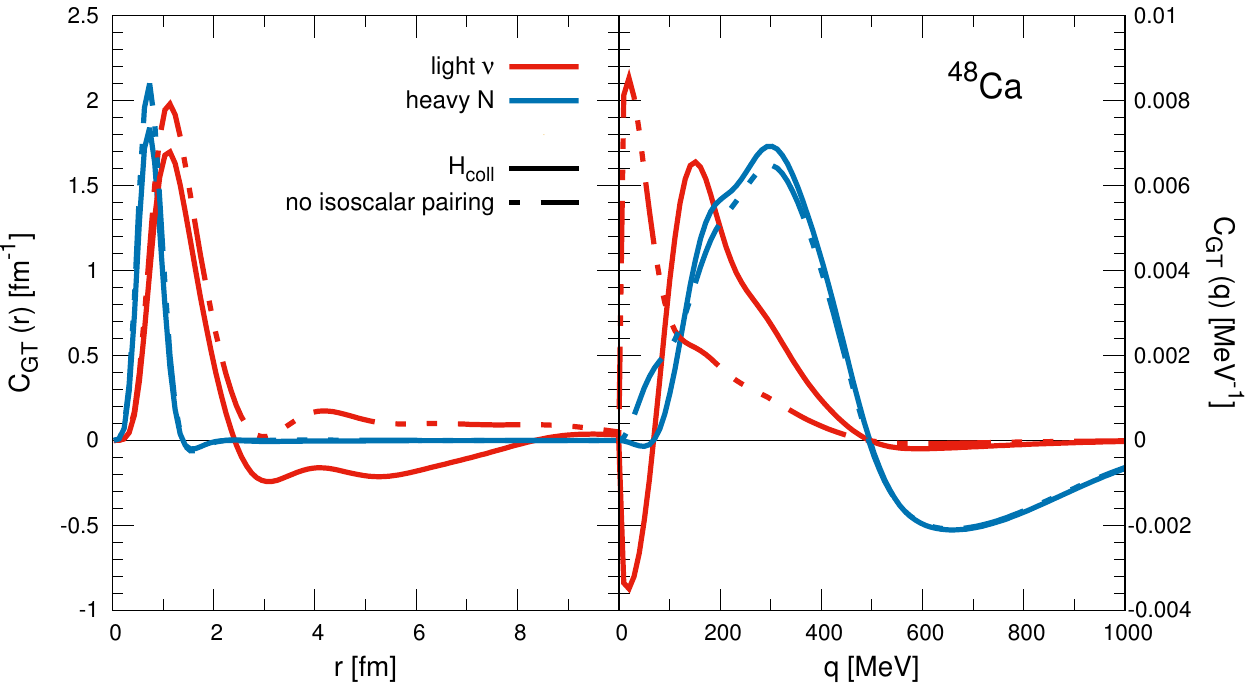}
\caption{Distribution of the Gamow-Teller matrix element of $^{48}$Ca.
Left panel: Radial distribution.
Right panel: Momentum transfer distribution.
Red (blue) lines represent the mechanism
that exchanges virtual light (heavy) neutrinos.
Solid (dashed) lines show results for the collective Hamiltonian
$H_{\mathrm{coll}}$ including (excluding) the isoscalar pairing term.
All results use Argonne-type short-range correlations.
The distributions are normalized with respect to $H_{\mathrm{coll}}$ results.}
\label{fig:nme_density_Hcoll}
\end{figure}

Figure~\ref{fig:nme_density_Hcoll} 
shows the radial and momentum transfer matrix element distributions,
$C(r_{ab})$ and $C(|\bm{q}|)$ in the two decay channels.
The only $\beta\beta$ emitter in the $pf$-shell,
$^{48}$Ca, is taken as an example.
We compare calculations performed using $H_{\mathrm{coll}}$
with and without the isoscalar pairing term.
The left panel of Fig.~\ref{fig:nme_density_Hcoll} shows that
for the exchange of light neutrinos the impact of isoscalar pairing
is significant at longer internucleon distances,
inducing matrix element cancellations beyond $3$~fm's or so.
When isoscalar pairing is included, the two decaying neutrons
can belong to different neutron-neutron and neutron-protons pairs,
making this interaction relevant at long distances~\cite{eng04}.
In comparison, there is only a mild reduction of the short-range
matrix element distribution when heavy neutrinos are exchanged.
Likewise, the right panel of Fig.~\ref{fig:nme_density_Hcoll} shows that
the variations in the momentum transfer distribution in the standard channel
are more marked for small momentum transfers,
where isoscalar pairing correlations reduce the value of the matrix element.
For heavy-neutrino exchange the momentum transfer distribution
is very much alike for the two Hamiltonians.

In conclusion, Fig.~\ref{fig:nme_density_Hcoll} highlights that
most of the contributions of isoscalar pairing affect
the long-range part of the $0\nu\beta\beta$ decay operator,
which is only important when light neutrinos are exchanged.
This implies a more pronounced role of isoscalar pairing correlations
in the standard light-neutrino-exchange scenario.
As originally motivated in Fig.~\ref{fig:nme_comp},
this is the channel where the variance between matrix elements is more marked.
In consequence, our analysis suggests that isoscalar pairing correlations
can be responsible for a sizeable part of the well-known disagreement
between shell-model and EDF theory $0\nu\beta\beta$ decay
nuclear matrix elements.

\begin{figure}
\includegraphics[width=\columnwidth]{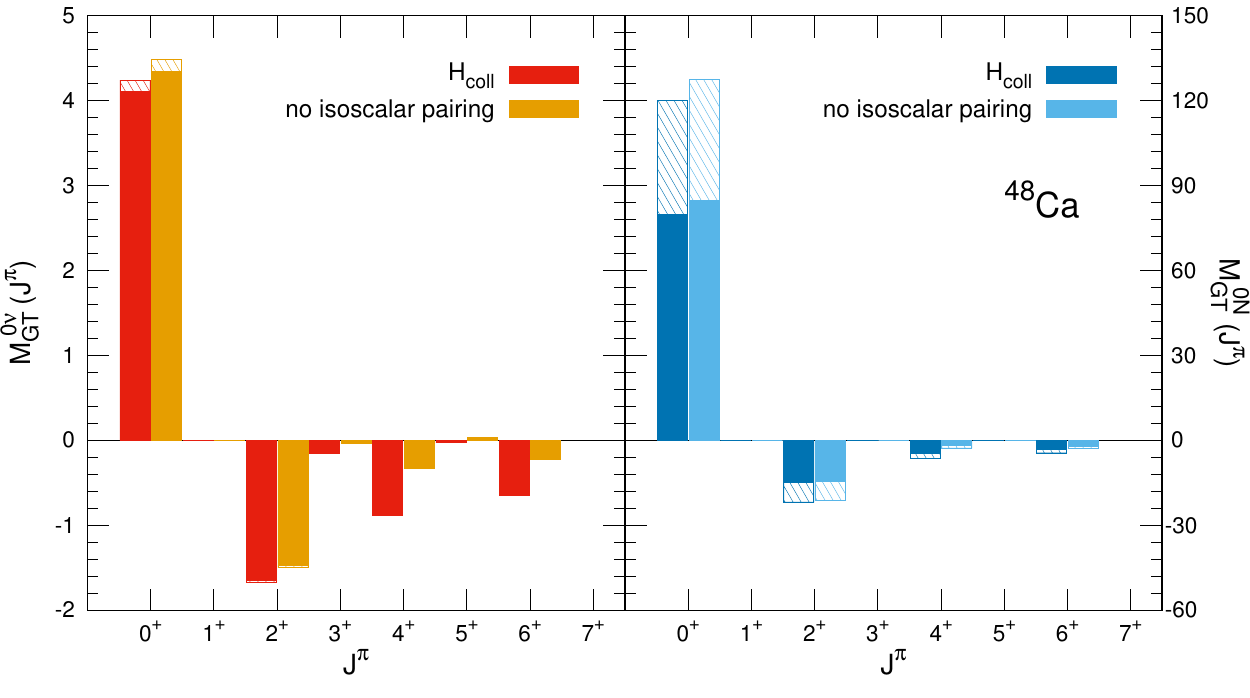}
\caption{Contributions to the Gamow-Teller matrix element of $^{48}$Ca
from pairs of decaying neutrons with given angular momentum and parity, $J^\pi$.
Left panel: Light-neutrino-exchange matrix elements, $M_{GT}^{0\nu}(J^{\pi})$.
Right panel: Heavy-neutrino-exchange matrix elements, $M_{GT}^{0N}(J^{\pi})$.
Red and blue (orange and light blue)
bars show results for $H_{\mathrm{coll}}$
($H_{\mathrm{coll}}$ without isoscalar pairing).
Solid (dashed) bars correspond to Argonne-type (CD-Bonn-type)
short-range correlations.}
\label{fig:nme_jp_Hcoll}
\end{figure}

Finally, Fig.~\ref{fig:nme_jp_Hcoll} shows
the angular momentum-parity decomposition of the $^{48}$Ca matrix element
using $H_{\mathrm{coll}}$, with and without the isoscalar pairing term.
When excluded, isoscalar pairing correlations are shown to impact especially
the contributions of $J>2$ pairs, which are much more significant
for the light-neutrino exchange shown in the left panel.
Isoscalar pairing correlations also reduce moderately
the $J=0^+$-pair contribution in both mechanisms,
but overall they are not very relevant
for heavy neutrinos, as the right panel shows.
This minor influence is consistent with the results in 
Figs.~\ref{fig:nme_Hcoll} and~\ref{fig:nme_density_Hcoll}.
The contributions of  $J>0$ pairs cancel the leading $J=0^+$ matrix element,
dominated by isovector pairing~\cite{caurier08}.
That the cancellation from $J>0$ pairs is driven by isoscalar pairing
supports that these correlations are important
to partially recover the $SU(4)$ symmetry of the nuclear states~\cite{men16},
because restoring the symmetry implies a suppression of the matrix elements.
The combined insights from
Figs.~\ref{fig:nme_density_Hcoll} and~\ref{fig:nme_jp_Hcoll}
suggest that $J>0$-pair matrix elements are dominated by
the longer-range part of the $0\nu\beta\beta$ decay operator,
which is particularly sensitive to isoscalar pairing correlations.

It is not straightforward how to extend the present analysis
beyond $pf$-shell nuclei,
because the shell model calculations of $\beta\beta$ emitters besides $^{48}$Ca
involve configuration spaces which miss spin-orbit partner orbitals,
and therefore are not suitable for the use of the collective Hamiltonian
in Eq.~(\ref{eq:hsep}).
Therefore the extent to which isoscalar pairing correlations
can impact the nuclear matrix elements of $^{76}$Ge or $^{136}$Xe is uncertain.
Nonetheless the estimates of Ref.~\cite{men16} suggest
that isoscalar pairing correlations are also relevant
for heavier $\beta\beta$ emitters,
although probably less so than in the extreme case,
illustrated for $^{48}$Ca in the right panel in Fig.~\ref{fig:nme_jp_Hcoll},
where the isoscalar pairing term is completely removed from $H_{\rm{coll}}$.
On the other hand, it is worth noting that very recently Ref.~\cite{Lay17}
has suggested, based on neutron-proton transfer reactions,
that a shell model interaction similar to the one used
to derive the isoscalar pairing strength of $H_{\rm{coll}}$
may underestimate the isoscalar pairing strength.


\section{Summary}
\label{sec:summary}

Nuclear matrix elements are key to determine the reach of
$0\nu\beta\beta$ decay searches, and to identify the decay channel
and obtain information of the underlying new physics
once $0\nu\beta\beta$ decay has been detected.
However, two of the preferred
many-body methods in nuclear structure, shell model and EDF theory,
predict matrix elements that differ by up to a factor three
in the standard scenario mediated by light-neutrino exchange.
In contrast, a similar comparison
for the matrix elements associated with the exchange of heavy neutrinos
shows a much milder disagreement of about $50\%$.
This reduced discrepancy is encouraging
toward a reliable determination of nuclear matrix elements.
In fact the difference between calculations is just about the size
of the uncertainty due to short-range correlations.

A comparison of the corresponding matrix elements
when correlations in the nuclear states are not permitted
shows an even better consistency between the shell model and EDF theory,
good to $30\%$ even for light-neutrino exchange.
This suggests that the variance between matrix elements
is mainly due to the unequal treatment by the many-body methods
of the nuclear structure correlations
that are hindered when heavy neutrinos are exchanged.
These correspond to the longer-range correlations
the $0\nu\beta\beta$ decay operator is sensitive to.
We have used a separable collective Hamiltonian to motivate that
the heavy-neutrino-exchange channel
is less sensitive to collective correlations.
In addition, we have observed that one of the main differences
between light- and heavy-neutrino exchanges is due to the role played by
isoscalar pairing correlations, which are mainly of long-range character
and contribute to cancellations in the matrix elements.
Our analysis supports previous works
suggesting that isoscalar pairing correlations can be responsible
for a sizeable part of the differences
between shell-model and EDF theory nuclear matrix elements
in the standard scenario of exchange of virtual light neutrinos.

Similar studies are needed to understand better the inconsistent
matrix elements obtained by various many-body approaches.
A more thorough investigation of quadrupole correlations,
not very relevant for the nuclei considered in our analysis,
is a natural extension of the present work.
Other possibilities include 
a comparison of the matrix elements corresponding to
alternative $0\nu\beta\beta$ decay channels
also considering the interacting boson model~\cite{bar15,isa17},
the QRPA~\cite{sim13,Hyvarinen15} and other novel methods,
focusing on the influence of additional collective correlations,
or testing the effect of extending the configuration space
---{\it e.g.} with perturbative~\cite{hol13c,Tsunoda16},
variational~\cite{Jiao17a,Jiao17b}
or improved shell-model~\cite{Shimizu17,Togashi16} techniques---
and other aspects of the many-body calculations.
Such studies can provide very valuable insights
to design strategies to improve matrix element calculations in the near future.

\section*{Acknowledgements}

J. M. would like to thank Y. Iwata for many estimulating exchanges
in the early stage of this work, 
V. Cirigliano, E. Mereghetti, T. Otsuka and N. Shimizu
for useful discussions,
and J. Engel for insightful comments on the manuscript.
This work was supported by 
the Japanese Ministry of Education, Culture, Sports, Science and Technology (MEXT)
as “Priority Issue on Post-K computer” (Elucidation of the Fundamental Laws and Evolution of the Universe) and 
the Joint Institute for Computational Fundamental Science (JICFuS).

\appendix
\section{Neutrino potentials
\label{sec:potentials}}
For completeness we give expressions for the neutrino potentials
defined in Sec.~\ref{sec:light_heavy},
used in the calculation of the nuclear matrix elements:
\begin{eqnarray}
h_{F}\left(|\bm{q}|\right)=h^{VV}_{F}\left(|\bm{q}|\right), \\
h_{GT}\left(|\bm{q}|\right)=h^{AA}_{GT}\left(|\bm{q}|\right)
+h^{AP}_{GT}\left(|\bm{q}|\right)+h^{PP}_{GT}\left(|\bm{q}|\right)
+h^{MM}_{GT}\left(|\bm{q}|\right), \\
h_{T}\left(|\bm{q}|\right)=h^{AP}_{T}\left(|\bm{q}|\right)
+h^{PP}_{T}\left(|\bm{q}|\right)+h^{MM}_{T}\left(|\bm{q}|\right),
\end{eqnarray}
where the superscripts indicate the particular terms
in the hadronic weak one-nucleon current
---vector, axial, pseudoscalar or magnetic--- generating each contribution.
Their explicit form is
\begin{eqnarray}
h^{VV}_{F}\left(|\bm{q}|\right) & = f_{V}^{2}\left(|\bm{q}|\right), \\
h^{AA}_{GT}\left(|\bm{q}|\right) & = f_{A}^{2}\left(|\bm{q}|\right), \\
h^{AP}_{GT}\left(|\bm{q}|\right) & = -\frac{2}{3} f_{A}^{2}\left(|\bm{q}|\right)
\frac{\bm{q}^{2}}{\left(\bm{q}^{2}+m_{\pi}^{2}\right)}\,, \\
h^{PP}_{GT}\left(|\bm{q}|\right) & = \frac{1}{3} f_{A}^{2}\left(|\bm{q}|\right)
\frac{\bm{q}^{4}}{\left(\bm{q}^{2}+m_{\pi}^{2}\right)^2}\,, \\
h^{MM}_{GT}\left(|\bm{q}|\right) & = \frac{2}{3}\left(\kappa_{p}-\kappa_{n}+1\right)^{2}\frac{g_V^2}{g_A^2}f_{V}^{2}\left(|\bm{q}|\right)
\frac{ \bm{q}^{2}}{4M_{N}^{2}}\,,  \\
h^{AP}_{T}\left(|\bm{q}|\right) & = \frac{2}{3} f_{A}^{2}\left(|\bm{q}|\right)
\frac{\bm{q}^{2}}{\left(\bm{q}^{2}+m_{\pi}^{2}\right)}\,, \\
h^{PP}_{T}\left(|\bm{q}|\right) & = -\frac{1}{3} f_{A}^{2}\left(|\bm{q}|\right)
\frac{\bm{q}^{4}}{\left(\bm{q}^{2}+m_{\pi}^{2}\right)^2}\,, \\
h^{MM}_{T}\left(|\bm{q}|\right) & = \frac{1}{3}\left(\kappa_{p}-\kappa_{n}+1\right)^{2}\frac{g_V^2}{g_A^2}f_{V}^{2}\left(|\bm{q}|\right)
\frac{ \bm{q}^{2}}{4M_{N}^{2}}\,,
\label{eq:h_q}
\end{eqnarray}
where $m_{\pi}\approx138$~MeV is the pion mass
and $\kappa_p-\kappa_n+1\approx4.70$~\cite{pdg14}
is the isovector nucleon magnetic moment.
The potentials $h^{AP}\left(|\bm{q}|\right)$
and $h^{PP}\left(|\bm{q}|\right)$
correspond to one- and two-pion-exchange diagrams, respectively.
The functions $f_{V}\left(|\bm{q}|\right)$ and $f_{A}\left(|\bm{q}|\right)$
reflect the structure of the nucleon, and are usually parameterized
with a dipole form factor:
\begin{eqnarray}
f_{V}\left(|\bm{q}|\right)=\frac{1}{\left(1+\bm{q}^2/\Lambda_V^2\right)^2}\,,
\quad
f_{A}\left(|\bm{q}|\right)=\frac{1}{\left(1+\bm{q}^2/\Lambda_A^2\right)^2}\,,
\end{eqnarray}
with $\Lambda_V=850$~MeV~\cite{Dumbrajs83}
and $\Lambda_A=1040$~MeV~\cite{Bernard01} taken from fits to experimental data.
At small momentum transfers ---not much larger than few times $m_{\pi}$---
equivalent expressions can be derived
in the context of chiral EFT~\cite{Bernard01}:
\begin{eqnarray}
f_{V}\left(|\bm{q}|\right)=1-2\frac{\bm{q}^2}{\Lambda_V^2}\,, \quad
f_{A}\left(|\bm{q}|\right)=1-2\frac{\bm{q}^2}{\Lambda_A^2}\,,
\end{eqnarray}
with these momentum-transfer corrections only entering at this order
in the neutrino potentials not suppressed
by $\bm{q}^{2}/M_{N}^{2}$~\cite{Menendez11}.
In the potentials $h^{AP}$ and $h^{PP}$
involving the pseudoscalar term in the hadronic current,
we assumed the Goldberger-Treiman relation for the pion decay constant
and pion-nucleon coupling, $f_{\pi}\,g_{\pi pn}\simeq g_A\,m_N$.
Corrections to this relation, which at small momentum transfers
are also predicted by chiral EFT,
are suppressed by $m_{\pi}^2/\Lambda_A^2$~\cite{sim09,Menendez11}.
In this work we neglect the additional contribution
of loop diagrams and counterterms predicted by chiral effective field theory~\cite{Cirigliano17c}.
First estimates suggest a $\sim 5\%$ effect for light-neutrino exchange,
and a $\sim10\%$ impact on heavy-neutrino-exchange nuclear matrix elements
for the isotopes considered here.

Table~\ref{tab:nme_decomposed} shows the decomposition of the
shell model $0\nu\beta\beta$ decay nuclear matrix elements
in Table~\ref{tab:nme} in terms of the contribution of each neutrino potential,
for light- and heavy-neutrino exchange.
\begin{table}[t]
	\caption{Contribution of each neutrino potential to the
		shell model $0\nu\beta\beta$ decay nuclear matrix elements for light- and heavy-neutrino exchange, $M^{0\nu}$ and $M^{0N}$. Left/right results denote Argonne-/CD-Bonn-type short-range correlations.}
	\begin{center}
	   \begin{tabular*}{\columnwidth}{lcccccc}
			\br
			&
			$^{48}$Ca & $^{76}$Ge & $^{82}$Se & $^{124}$Sn & $^{130}$Te & $^{136}$Xe \\
			\hline \\[-1em]
			$M^{0\nu}$ &
			0.93/1.02 & 2.89/3.07 & 2.73/2.90 & 2.76/2.96 & 2.76/2.96 & 2.28/2.45  \\ \\[-1em]
			$\hspace{-0.3cm}-M_{F}^{VV}$ &
			0.23/0.25 & 0.54/0.59 & 0.51/0.55 & 0.61/0.66 & 0.62/0.67 & 0.50/0.54 \\ \\[-1em]
			$M_{GT}^{AA}$ &
			1.00/1.08 & 2.98/3.15 & 2.81/2.97 & 2.79/ 2.97 & 2.78/2.97 & 2.31/2.45 \\ \\[-1em]
			$\hspace{-0.3cm}-M_{GT}^{AP}$ &
			0.35/0.38 & 0.87/0.94 & 0.82/0.89 & 0.87/0.95 & 0.89/0.97 & 0.73/0.79 \\ \\[-1em]
			$M_{GT}^{PP}$ &
			0.12/0.13 & 0.27/0.30 & 0.25/0.28 & 0.27/0.30 & 0.28/0.31 & 0.23/0.25 \\ \\[-1em]		
			$M_{GT}^{MM}$ &
			0.08/0.10 & 0.18/0.22 & 0.17/0.20 & 0.19/0.23 & 0.19/0.23 & 0.16/0.19 \\ \\[-1em]	
			$M_{T}^{AP}$ &
			-0.08/-0.08 & -0.01/-0.01 & -0.01/-0.01 & 0.01/0.01 & 0.01/0.01 & 0.01/0.01 \\ \\[-1em]
			$M_{T}^{PP}$ &
			0.03/0.03 & 0.00/0.00 & 0.00/0.00 & -0.01/-0.01/ & -0.01/-0.01 & -0.01/-0.01 \\ \\[-1em]		
			$M_{T}^{MM}$ &
			-0.01/-0.01 & 0.00/0.00 & 0.00/0.00 & 0.00/0.00 & 0.00/0.00 & 0.00/0.00 \\			  			  			  
			\mr
			$M^{0N}$ &
			48/73 & 130/188 & 121/175 & 141/203 & 146/210 & 116/167 \\ \\[-1em]
			$\hspace{-0.3cm}-M_{F}^{VV}$ &
			21/26 & 49/59 & 46/55 & 52/63 & 54/65 & 43/52 \\ \\[-1em]
			$M_{GT}^{AA}$ &
			65/85 & 150/196 & 140/183 & 158/208 & 163/214 & 130/171 \\ \\[-1em]
			$\hspace{-0.3cm}-M_{GT}^{AP}$ &
			29/42 & 63/91 & 59/84 & 68/98 & 70/101 & 56/80 \\ \\[-1em]
			$M_{GT}^{PP}$ &
			10/16 & 21/33 & 19/31 & 22/36 & 23/37 & 18/30 \\ \\[-1em]			
			$M_{GT}^{MM}$ &
			-2/7 & -6/14 & -6/13 & -5/17 & -5/17 & -4/14 \\ \\[-1em]		
			$M_{T}^{AP}$ &
			-12/-12 & -2/-2 & -2/-2 & 2/2 & 3/3 & 2/2 \\ \\[-1em]
			$M_{T}^{PP}$ &
			5/5 & 1/1 & 1/1 & -1/-1 & -1/-1 & -1/-1 \\ \\[-1em]			
			$M_{T}^{MM}$ &
			-2/-2 & 0/0 & 0/0 & 0/0 & 0/0 & 0/0 \\ \\[-1em]
			\br
		\end{tabular*}
	\end{center}
	\label{tab:nme_decomposed}
\end{table}

\section*{References}

\bibliographystyle{iopart-num}
\bibliography{bb-ropp}

\end{document}